\documentclass[12pt]{iopart}

\usepackage{graphics}
\usepackage{graphicx}
\begin{document}

\title[Dorso, Gim\'enez Molinelli, L\'opez]{Isoscaling and the nuclear EOS}

\author{C.A. Dorso$^1$, P.A. Gim\'enez Molinelli$^1$, and
J.A. L\'opez$^2$}

\address{$^1$ Universidad de Buenos Aires, Buenos Aires,
Argentina}
\address{$^2$ University of Texas at El Paso, El Paso, Texas 79968, USA}

\ead{jorgelopez@utep.edu}

\begin{abstract}
Experiments with rare isotopes are shedding light on the role isospin plays in the equation of state (EoS) of nuclear matter, and isoscaling --an straight-forward comparison of reactions with different isospin-- could deliver valuable information about it.  In this work we test this assertion pragmatically by comparing molecular dynamics simulations of isoscaling reactions using different equations of state and looking for changes in the isoscaling parameters; to explore the possibility of isoscaling carrying information from the hot-and-dense stage of the reaction, we perform our study in confined and expanding systems.  Our results indicate that indeed isoscaling can help us learn about the nuclear EoS, but only in some range of excitation energies.
\end{abstract}

\maketitle

\section{Motivation}
\label{intro}

Experimental advances in the last decade have allowed the manufacturing of rare isotopes --unstable nuclei with excess neutrons-- in ways suitable for reactions, this has moved the frontier of nuclear science in the dimension of isospin. First results coming from studies of rare isotopes shattered the immutability of the shell structure~\cite{Rah10} and magic numbers and drew a road map to use neutron-rich nuclei to study the elusive drip lines which lie far from stable isotopes~\cite{NSA06}.  These and other results have propelled the isotopic degree of freedom to the spotlight, and new facilities and experiments promise to help us obtain an improved description of nuclear masses~\cite{Dan02}, fission barriers, collective vibrations~\cite{Kli07}, neutron skins of neutron-rich nuclei \cite{Bro00} and, in general, understand the equation of state (EoS) for neutron-rich matter.

To reap all these benefits, however, the effect that varying isospin has on the EoS must be known and, up to date, this is not the case.  Information to constrain the nuclear EoS at non-saturation densities and non-zero temperatures has been obtained from heavy ion reactions.  These collisions produce, albeit for a brief moment, nuclear systems at high density and temperature and thus serve as probes of characteristics of the sought EoS.  This is mainly achieved through comparisons of experimental observables to theoretical calculations; this program, however, faces a series of challenges both through the detection of clear experimental signatures and the crafting of reliable theoretical simulations.

The main challenge comes from trying to use an EoS to describe a heavy ion reaction. In a perfect world, in which the nucleon-nucleon interacting potentials were known, excited nuclei in reactions could be described in terms of these interactions. Out of these microscopic descriptions of nuclear processes, one could then use appropriate averaging procedures to obtain macroscopic information, (e.g. $\rho$, $\epsilon$, $K$, $T$, $p$, mean fields, enthalpy, etc.) which could be used to craft an EoS.  A major condition for this plan to work is the attainment of thermal and chemical equilibrium as the macro variables (and the EoS) lose meaning in non-equilibrium processes; it is clear that the micro description (in a perfect world) would be appropriate to describe all the in- and out-of-equilibrium stages of a reaction while the EoS is not.  Thus a reaction cannot be fully described in terms of the EoS or macro variables, and information about the EoS can only be obtained from those segments of the reaction in which the macro variables stabilize; this can only be accomplished by means of careful comparison of computational simulations of reactions to experimental data.

To understand the role isospin plays in the EoS, one must first find an observable that can provide valuable experimental information that could be analyzed within a sensible theoretical framework; initial studies have attempted to do that through isoscaling.  When comparing the fragmentation yield of stable nuclei reactions with those of neutron-rich nuclei collisions, it was observed that they are related by  a power law of the form $Y_2(N,Z) \propto Y_1(N,Z)e^{-\alpha N + \beta Z}$, where $N$ and $Z$ are the neutron and proton content of the produced fragments, and $\alpha$ and $\beta$ are fitting parameters~\cite{xu,tsang2001}.

The theoretical framework first used to interpret the fitting parameters was that of hot and dense {\em fragmenting source} disassembling into fragments while equilibrated in micro or grand canonical~\cite{tsang}, or canonical~\cite{das} ensembles.  Due to a lack of a better approach, the isospin dependence of the binding energy of cold nuclei at normal saturation density was assumed to be valid at other densities and higher temperatures to find an interpretation of the isoscaling parameters. In this approximation the isoscaling parameters were found to be connected to the symmetry energy term of the Weissacker mass formula, $E_{sym}=C_{sym}(A-2Z)^2/A$ through, $\alpha=4C_{sym}[(Z_1/A_1)^2-(Z_2/A_2)^2]/T$, where the subscripts refer to the proton and nucleon numbers of the fragmenting sources created in the two reactions \cite{tsang2001, ono}.

Unfortunately, the lack of density or temperature dependence of the mass formula, as well as other problems, blur the final interpretation of $\alpha$ and $\beta$.  A major concern is the fact that $T$ and $\rho$ vary during the reaction, and the concept of a {\em fragmentation source} is only an approximation useful for invoking  statistical models; indeed, the validity of $C_{sym}$ (and thus the interpretation of $\alpha$) have been strongly questioned~\cite{Bao06}.  Along the same lines, it is known that the isoscaling parameters are know to vary widely during the reaction~\cite{Bar07}; this casts further doubts about the proposed interpretation of $\alpha$.

Thus we arrive at the motivation of the present study: what can we learn about the isospin dependence of the EoS from isoscaling?  Does isoscaling reflect properties of a {\em fragmentation source}, or is it set during the expansion of the disassembling source?  Can isoscaling tell us something about, say, the compressibility of nuclear matter?  These and other questions will be answered in this study.

\subsection{Plan of action}\label{plan}
Since the process of extracting the isoscaling parameters from experiments is very entangled requiring averages and fits over hundreds of different collisions, we feel it would be futile to try to connect the parameters to a possible $\rho$ and $T$-dependent symmetry term, $C_{sym}(\rho,T)$.  Unfortunately, other promising approaches, such as those extending the EoS to the non-symmetric case through, e.g., $E(\rho,\delta)=E(\rho,\delta=0)+S(\rho)\delta^2$, where the isospin is inputted through the neutron, proton and nucleon densities, $\delta=(\rho_n-\rho_p)/\rho$~\cite{Tsa09}, have not yet been linked to isoscaling.  In consequence, we opt to follow a much more pragmatic plan of action to understand the connection of isoscaling to the EoS.

In brief, we perform simulations of isoscaling reactions using different equations of state and look for any differences in the behavior of the isoscaling parameters; same results from different EoS would practically nullify isoscaling as a probe, different behavior under different EoS would hold promise.  Furthermore, to explore the possibility (implied by the proponents of the relationship between $\alpha$ and $C_{sym}$) that isoscaling carries information from the hot-and-dense stage of the reaction to asymptotia, we carry out the proposed study both for systems disassembling under confined conditions and during expansion; again, different results would indicate that isoscaling is affected by the expansion, else it would hold promise as a probe of the fragmentation source.

In section~\ref{sec1} we present argument to justify the computational model used in this study, describe how the symmetry energy terms were obtained for each of the parameterizations of the potential, and obtain their caloric curves. In section~\ref{sec2} we review the concept of symmentropy and study the isoscaling of confined and expanding systems. The manuscript closes with a summary of the main conclusions.

\section{Model}\label{sec1}
In recent articles we have studied the behavior and properties of isoscaling using both classical and geometrical models. In \cite{dorso_pandha} we found isoscaling in simulations of classical systems and in~\cite{dorso_perco} we have shown, both analytically and numerically, that isoscaling can also be observed in the framework of the nuclear percolation model; an effect totally due to the probabilistic aspects of the problem.  Along the same line of work, in~\cite{dorso} we studied the effect particle correlations have on the isoscaling parameters, and in~\cite{dorso_moretto} we finally concluded that a minimum isoscaling can be expected for any disassembling system based solely on probabilistic aspects.

In this article we go back to using classical molecular dynamics to study isoscaling produced in systems with two different EOS and disassembling in confined and expanding environments.  In particular, we will use two parameterizations of the interaction potential~\cite{pandha} that lead to EoS with compressibility of $250 \ MeV$ and $530 \ MeV$, and will study equilibrated systems clustering in a spherical container as well as in a free expansion.

\subsection{Molecular dynamics model}
Here the molecular dynamics model used for this study is introduced along with the two potentials that lead to different compressibility values.  To study the origin of isoscaling, a model capable of reproducing both the out-of-equilibrium and the equilibrium parts of a collision is highly desirable; in the present work, we use a molecular dynamics ($MD$) model that can describe non-equilibrium dynamics, hydrodynamic flow and changes of phase without adjustable parameters.  The combination of this $MD$ code with a fragment-recognition algorithm has been applied in recent years to study, among other phenomena, neck fragmentation~\cite{chernolaval}, phase transitions~\cite{oax2001}, and other features of nuclear reactions, including isoscaling~\cite{HIP2003,dorso_pandha}.

The $MD$ code uses a two-body potential composed of the Coulomb interaction plus a nuclear part composed of two Yukawa-like potentials (known as the Illinois potential ~\cite{pandha}) that correctly reproduces nucleon-nucleon cross sections, as well as the correct binding energies and densities of real nuclei.
The \textquotedblleft nuclear\textquotedblright\ part of the interaction potential is
\begin{eqnarray*}
V_{np}(r) =&V_{r}\left[ exp(-\mu _{r}r)/{r}-exp(-\mu _{r}r_{c})/{r_{c}}%
\right] \\
&\ \mbox{}-V_{a}\left[ exp(-\mu _{a}r)/{r}-exp(-\mu _{a}r_{a})/{r_{a}}%
\right] \\
V_{nn}(r)&=V_{pp}(r)=V_{0}\left[ exp(-\mu _{0}r)/{r}-exp(-\mu _{0}r_{c})/{%
r_{c}}\right]  \label{2BP}
\end{eqnarray*}
where the cutoff radius is $r_{c}=5.4~fm$, $V_{np}$ is the potential between a neutron and a proton while $V_{nn}$ is that between identical nucleons. The values of the parameters of the Yukawa potentials can be selected~\cite{pandha} as to correspond to an equation of state of infinite nuclear matter with an equilibrium density of $\rho_{0}=0.16~fm^{-3}$, a binding energy $E(\rho _{0})=-16~MeV/nucleon$, and a compressibility of $\sim 250$ $MeV$ for the so called medium model, or of $\sim 535$ $MeV$ for the stiff model.

\subsection{Symmetry energy coefficients}
For completeness, we start by verifying that these two types of interactions indeed lead to different values of the symmetry energy for cold nuclei at saturation density.  To achieve this, we first used dissipative molecular dynamics to construct ``nuclei'' of several masses in their ground states, and then fit the results with the mass formula thus obtaining values of the different coefficients, including the symmetry energy term, $C_{sym}$.

\begin{figure}
\centerline{\includegraphics[scale=0.5,angle=-0]{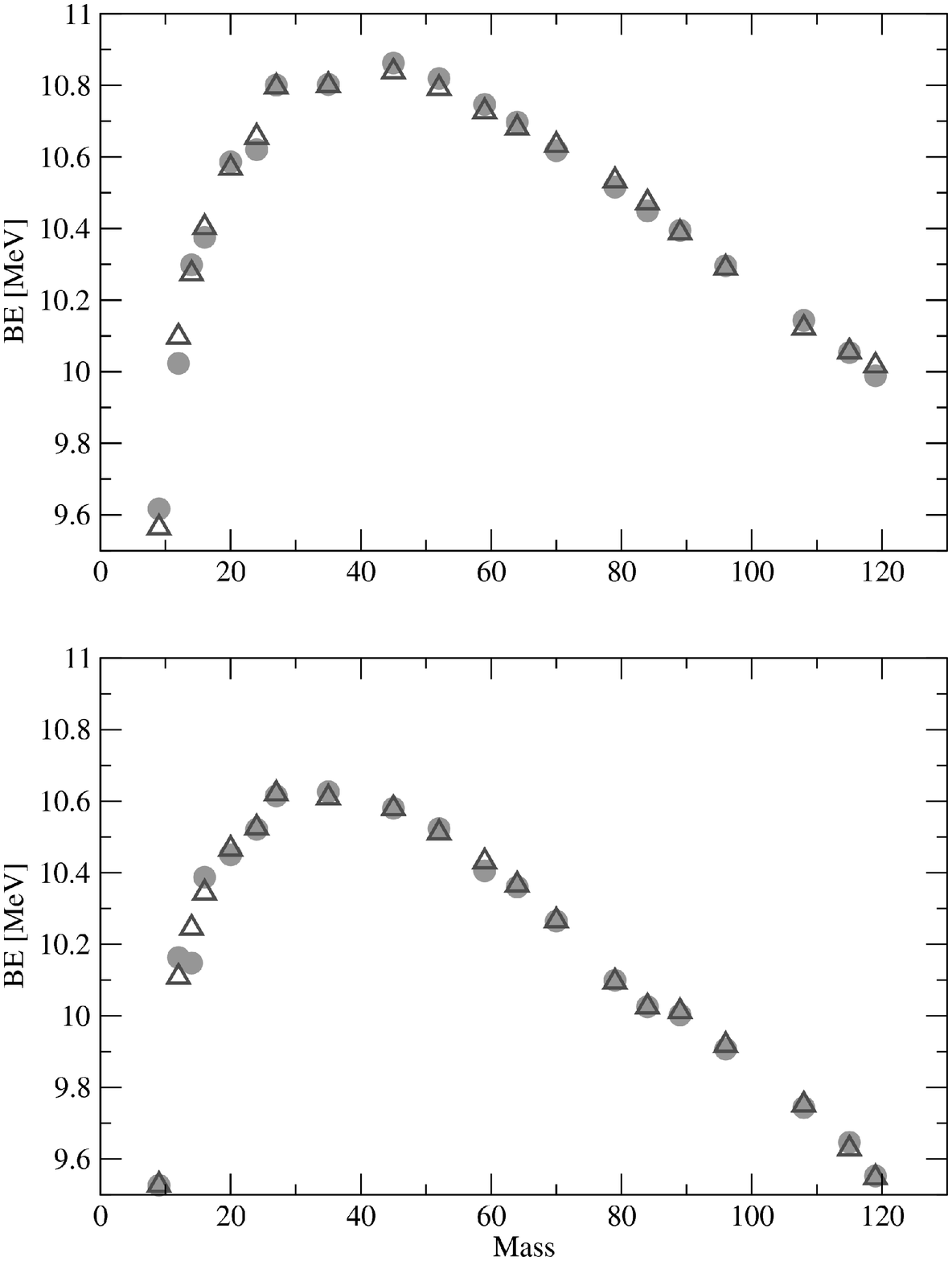}}
\caption{Energies obtained with the mass formula fit (triangles)
for the Stiff and Medium models (top and bottom panels respectively)
together with the corresponding ground states calculated using
frictional molecular dynamics (circles)} \label{fig1}
\end{figure}

In a nutshell, in dissipative MD individual nuclei are evolved and cooled until they become self-bound and reach proper values of the nuclear radius and binding energy.  Taking such state as the ground state (still with some residual motion which mimics Fermi motion) the operation is repeated for many nuclei of different sizes.  We then proceeded to fit the resulting nuclei with the liquid-drop mass formula for the nuclear binding energy:
\begin{equation*}
E/A=C_{v}+C_{s} A^{-1/3}+4 C_{c} \left( Z\left( Z-1\right)
\right) A^{-4/3}\\+C_{sym}(N-Z)^{2}A^{-2},
\end{equation*}
\noindent thus obtaining the values of the coefficients presented in Table~\ref{tab1} alongside with published experimental data~\cite{rohlf} for comparison.  In Table \ref{tab2} we list the nuclei used to fit the binding energy, and in Fig.~\ref{fig1} we show the binding energies and droplet model values obtained for both medium and stiff models.

\begin{table}
 \centering
 \begin{center}
\caption{Comparison of coefficients obtained for different models}
\label{tab1}       
\begin{tabular}{llll}
\hline\noalign{\smallskip}
Coefficient & Stiff & Medium & Experimental  \\
\noalign{\smallskip}\hline\noalign{\smallskip}
$C_{v}$ & 16.1 & 17.37 & 15.75 \\
$C_{s}$ & -11.73 & -14.38 & -17.8 \\
$C_{c}$ & -0.197 & -0.226 & -0.177\\
$C_{sym}$ & -34.07 & -25.08 & -23.7\\
\noalign{\smallskip}\hline
\end{tabular}
 \end{center}
\end{table}

\begin{table}
 \centering
 \begin{center}
\caption{Nuclei used to fit binding energy formula coefficients}
\label{tab2}       
\begin{tabular}{lllll}
\hline\noalign{\smallskip}
$A$ & $N$ & $Z$ & $E_{Med}$ & $E_{Stiff}$ \\
\noalign{\smallskip}\hline\noalign{\smallskip}
  9 &  5 &  4 &  9.527 &  9.617 \\
 12 &  6 &  6 & 10.163 & 10.023 \\
 14 &  7 &  7 & 10.148 & 10.299 \\
 16 &  8 &  8 & 10.387 & 10.376 \\
 20 & 10 & 10 & 10.450 & 10.586 \\
 24 & 12 & 12 & 10.521 & 10.620 \\
 27 & 14 & 13 & 10.614 & 10.801 \\
 35 & 18 & 17 & 10.660 & 10.803 \\
 45 & 24 & 21 & 10.580 & 10.862 \\
 52 & 28 & 24 & 10.523 & 10.819 \\
 59 & 32 & 27 & 10.405 & 10.747 \\
 64 & 35 & 29 & 10.360 & 10.697 \\
 70 & 39 & 31 & 10.265 & 10.618 \\
 79 & 45 & 34 & 10.100 & 10.516 \\
 84 & 48 & 36 & 10.026 & 10.449 \\
 89 & 50 & 39 & 10.002 & 10.395 \\
 96 & 54 & 42 &  9.907 & 10.296 \\
108 & 61 & 47 &  9.744 & 10.143 \\
115 & 66 & 49 &  9.647 & 10.054 \\
119 & 69 & 50 &  9.553 &  9.989 \\
\noalign{\smallskip}\hline
\end{tabular}
\end{center}
\end{table}

These reassuring results indicate that indeed the model used is capable of mimicking nuclei with different compressibilities, i.e. with different symmetry energy coefficients.  Consequently, if isoscaling is to vary under different equations of state, the model most likely will show such differences if used with the two parametrizations of the potential.  We now proceed to use the model to detect differences in isoscaling due to different values of the symmetry energy.

\subsection{Simulating the disassembly}
In previous works we have analyzed, among others, the collisions of ${}^{40}$Ca+${}^{40}$Ca, ${}^{48}$Ca+${}^{48}$Ca and ${}^{56}$Ca+${}^{56}$Ca at different energies, in this work we start by studying the evolution of constrained ``nuclei'' of $(N,Z)$ $ =(40,40)$ and $(56,40)$, which correspond to the complete merging of the colliding nuclei of the previous works.

As done above we have used the dissipative molecular dynamics method to build nuclei in their ground states. Once the ground states are available, we place the resulting cold systems inside a spherical container. The radii of the container is fixed in order to obtain selected values of the number density ranging from $0.001~fm^{-3}$ to $0.007~fm^{-3}$. To produce the disassembly, energy is added by scaling the momenta of the particles. The trajectories of motion of individual nucleons are then calculated using the standard Verlet algorithm with an energy conservation of $\mathcal{O}$($0.01\%)$. We first perform a long run in order to let the system relax to equilibrium. Afterwards a much longer run is performed and well separated in time snapshots of the evolution are recorded.

From the microscopic information of the evolution, given by the values of position and momenta of the nucleons, we calculate the fragment structure of the system by means of the MSTE cluster-detection algorithm introduced decades ago~\cite{hill}, but recently adapted for the nuclear case~\cite{mse}. According to this prescription, a particle $i$ belongs to a cluster $C$ if there is a particle $j$ in $C$ to which $i$ is bound in the sense of $p_{ij}^{2}/{4\mu } <v_{ij}$, where $p_{ij}$ is the magnitude of the relative momentum, $\mu $ the reduced mass, and $v_{ij}$ the interparticle potential. In this cluster definition the effect of the relative momentum between the particles that form the cluster is taken into account in an approximate way. It should be kept in mind
that these clusters correspond to a given snapshot of the evolution. They are not stable in the sense that their lifetime is a function of the particle-particle collision frequency in the system.

The systems $(N,Z)=(40,40)$ and $(56,40)$, were studied at different energies in the range $-5$ to $8\ MeV/A$ and for four values of the number density with two thousand snapshots recorded at each energy.  It must be remarked that the $MD$ model is fully classical and all quantal effects, such as the exclusion principle, Fermi motion and isotopic content-modifying phenomena, are excluded. Therefore, any observed variations to isoscaling will be entirely due to the change of EOS.  [The effect of the Fermi motion, although formally absent, is somewhat included by the internal motion of the ``nucleons'' in their ``ground state'' as explained before.]

\subsubsection{Caloric curves}
\begin{figure}
\centerline{ \includegraphics[scale=0.40,angle=-90]{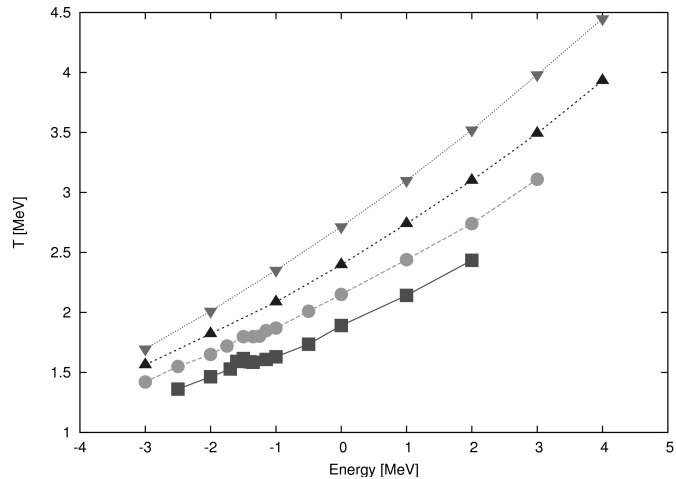}}
\caption{Caloric curves for the (40,40) system with the
medium model for several number densities. Squares are for
$0.001~fm^{-3}$, circles for $0.003~fm^{-3}$,
up triangles for $0.005~fm^{-3}$ and down triangles for
$0.007~fm^{-3}$} \label{fig2}
\end{figure}

Before proceeding to the study of isoscaling we first characterize the disassembling source by its caloric curve, i.e. the functional relationship between the temperature and energy of the system.  Calculating the temperatures through the usual $T=(2/3N)\sum p_{i}^{2}/2m$, with $p_{i}$ representing the momentum of particle $i$, and the sum runs through the $N$ particles in the system, the caloric curves for the medium model are displayed in Fig.~\ref{fig2} as a function of the energy of the system and for several densitites.

It is interesting to notice that the systems displays a behavior already found in Lennard Jones systems~\cite{ortiz,ison_coulomb}, i.e. as the density is lowered the caloric curves start to develop a ``loop'' which signals the presence of a negative specific heat. This behavior has been traced to the appearance of ``surfaces'' in the system, and has been associated with a first order phase transition.

\section{Isoscaling}\label{sec2}

\subsection{Symmentropic limit}
It has been shown that isoscaling is an effect generic to disassemblying systems which can be characterized by the concept of symmentropy~\cite{dorso_moretto, dorso}; as this is used in  the following analysis, we first introduce it here.  In summary, and for the nuclear case, the comparison of two percolating lattices ($i=1,2$) with nodes occupied by neutrons and protons with occupation probabilities $p_{Z_i}=Z_i/A_i$ and $p_{N_i}=N_i/A_i$, but similar bond-breaking probabilities, produce the ratio of yields $R_{21}(N,Z)={Y_{2}(N,Z)}/{Y_{1}(N,Z)}=
\left[ {p_{Z_{2}}}/{p_{Z_{1}}}\right]^{Z}\left[ {
p_{N_{2}}}/{p_{N_{1}}}\right]^{N}$, which is directly related to the isoscaling power law $R_{21}(N,Z) \propto e^{(\alpha N + \beta Z)}$ with $\alpha =\ln (p_{N_{2}}/
p_{N_{1}})$ and $\beta =\ln (p_{Z_{2}}/p_{Z_{1}})$; we take these as the symmentropic limit of the isoscaling parameters to be used for comparison.

As an example, for isoscaling between $^{40}$Ca+$^{40}$Ca and  ${}^{48}$Ca+$^{48}$Ca, we have $A_1=80$, $N_1=40$, $A_2=96$, $N_2=56$, and thus $p_{N_1}=40/80$, $p_{N_2}=56/96$, and $\alpha=\ln(1.16)=0.154$; $\alpha$'s near this limit will carry only probabilistic information and will not allow us to use isoscaling to probe the EoS.

\subsection{Isoscaling for confined systems}\label{isoscaling}

Data from the evolutions of confined systems as produced in the reactions $^{40}$Ca+$^{40}$Ca and ${}^{48}$Ca+$^{48}$Ca were used to construct the yield matrices $Y_{i}(N,Z)$ for the two reactions ($i=1,2$) and the corresponding matrix $R_{21}(N,Z)=Y_{2}(N,Z)/Y_{1}(N,Z)$.  Least squares fits to the isoscaling power law yielded values of the parameters $\alpha $ and $\beta $ for each of the snapshots of the evolution at the conditions of density and excitation energy stated above.

\begin{figure}
\centerline{ \includegraphics[scale=0.45,angle=-90]{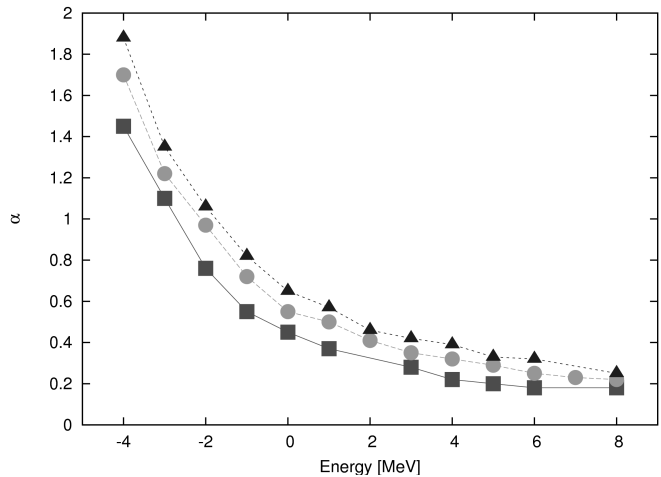}}
\caption{Values of isoscaling coefficient $\alpha$ for the medium
model. Squares for number density $0.001~fm^{-3}$, circles for
$0.003~fm^{-3}$ and triangles for $0.007~fm^{-3}$.} \label{fig3}
\end{figure}

Fig.~\ref{fig3} shows the behavior of $\alpha$ as a function of excitation energy for different densities of the medium model; similar results were obtained for the stiff model.  It is worth mentioning that $\alpha $ is a decreasing function of the energy of the system, but more important, it converges to a constant value with asymptotically appears to approach the symmentropic limit; this suggests that at high energies the breakup resembles more and more a simple probabilistic disassembly.

\begin{figure}[htb]
\centerline{ \includegraphics[scale=0.6,angle=-90]{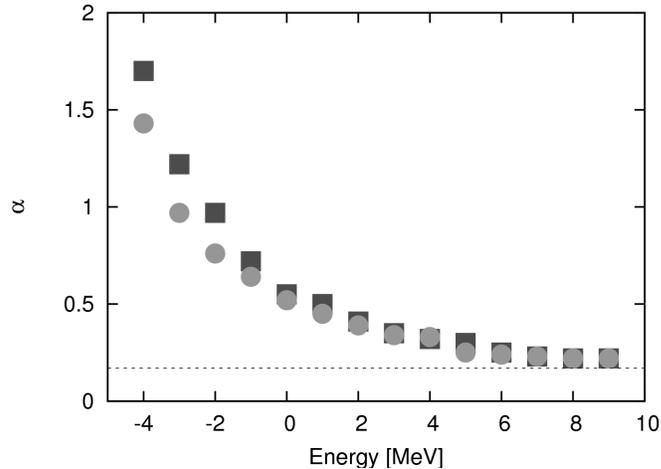}}
\caption{$\alpha$ as a function of energy obtained from confined
system for the two models at a fixed value of the density
$\rho=0.007~fm^{-3}$. Squares for the medium model and circles
for the stiff; the dashed line shows the symmentropic limit.} \label{fig4}
\end{figure}

Repeating this type of simulations and analyses for the two interaction potentials, can allow us to explore the use of isoscaling to differentiate between a medium-compressibility and stiff-compressibility equations of state.  For this purpose we have fixed the number density of our confined systems to $N/V=0.007~fm^{-3}$ and performed the same calculations for both potentials; the resulting values of $\alpha$ are displayed in Fig.~\ref{fig4}.

Although the medium model has values of $\alpha$ which exceed those of the stiff model by as much as $\sim 15\%$, this difference appears to be restricted to the low (self-bound) energy regime.  At high energies, both compressibilities appear to have values of $\alpha$ that are indistinguishable from one another and both converging to the symmentropic limit at high energies (see dashed line).

\subsection{Isoscaling for free expanding systems}
To assess the effect of the expansion on isoscaling, we now reproduce the previous calculations for systems that underwent an expansion.  For this purpose we take the configurations recorded for the confined equilibrated systems, remove the confining walls, and follow the subsequent evolution by solving the equations of motion.  Once that the produced fragments and particles are separated and de-excited enough, the steps leading to the $\alpha$ parameter are repeated on these asymptotic configurations.

\begin{figure}[htb]
\centerline{ \includegraphics[scale=0.6,angle=-90]{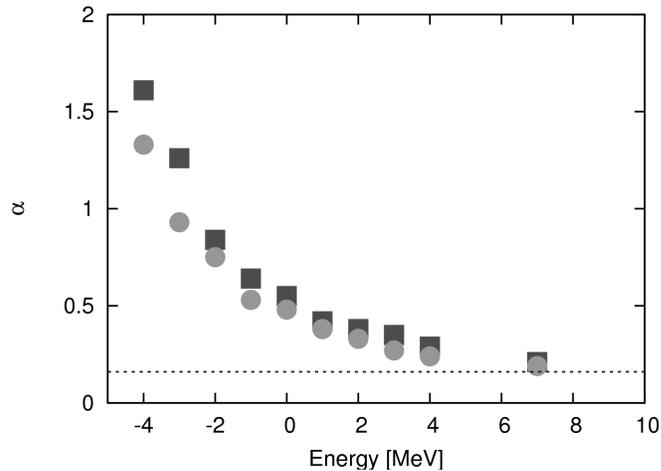}}
\caption{Asymptotic values of $\alpha$ obtained from systems expanding from initial densities of $\rho=0.007~fm^{-3}$ as a function of their excitation energy.  Squares for the medium model and circles for the stiff; the dashed line shows the symmentropic limit.} \label{fig5}
\end{figure}

Fig.~\ref{fig5} shows the values of the $\alpha$'s obtained from these runs. Notice that each point in this plot corresponds to a point in Fig.~\ref{fig4}, and thus allow us to detect a small decrease of the overall values of $\alpha$ from confinement to expansion.  The asymptotic values of the isoscaling parameters indeed are a bit smaller but comparable to those generated at the fragmentation stage of the reaction.  Therefore,  isoscaling appears to have the intrinsic ability to provide information of the early moments of the reaction when, presumably, the nuclear system is in equilibrium and the information reflects properties of the nuclear EoS.

And that was the good news, the not-so-good- news is that, as found before for confined systems, isoscaling can help us differentiate between stiff and medium compressibility EoS only at low energies.  Again, systems expanded at high energies tend to isoscaling values resembling those predicted by symmentropy and have no EoS information to give us.

\section{Conclusions}

To quantify the role of isoscaling to provide us with information from the nuclear EoS, we have studied isoscaling using molecular dynamic simulations.  To assess the relay of information from the hot-and-dense stage of the reaction to asymptotia, the study was performed with confined systems and compared to similar systems after undergoing an expansion.  To measure the sensibility of the isocaling parameters to the EoS, the study was performed with two systems corresponding to different compressibilities.  To further connect these calculations with the Weissacker mass formula, the symmetry energy coefficient of these medium and stiff media were calculated directly from dissipative simulations.

The disassembling systems studied were equivalent to those formed by collisions of $^{40}$Ca+$^{40}$Ca and  ${}^{48}$Ca+$^{48}$Ca, and with total binding energies ranging from $-4$ to $8$ MeV/A. The resulting fragment data were used to construct the ratios $R_{21}(N,Z)$ and to obtain the fitting parameters $\alpha$ and $\beta$ of the isoscaling power law; the analysis, though, focused on $\alpha$.

In summary, studies of both confined and expanded systems and with the different interaction potentials demonstrated that:
\begin{itemize}
\item[$i)$] Confined systems produce slightly larger, but comparable, values of $\alpha$ than expanded systems at low energies; the difference disappears at higher energies.
\item[$ii)$] In both confined and expanded cases, the medium-compressibility system produces slightly larger $\alpha$ than the stiff one, again, this difference is restricted to the low excitation energy regime,
\item[$iii)$] in all cases the values of $\alpha$ decrease with increasing excitation energy converging asymptotically to the symmentropic limit in agreement to previous studies~\cite{isosc_high_energy}.
\end{itemize}

\noindent These results strongly suggests that:
 \begin{itemize}
\item[$iv)$] isoscaling indeed has the ability to carry information from the early moments of the collision, i.e. an asymptotic value of $\alpha$ can be taken as representative of the conditions in the hot-and-dense stage of the reaction.
\item[$v)$] Likewise, the approach to the symmentopic limit, seems to indicate that at high energies the breakup approaches a simple probabilistic disassembly; $\alpha$ values at these energies do not carry any nuclear information.
\item[$vi)$] Finally, the use of isoscaling to distinguish between stiff and medium-compressibility EoS must be limited to low energies.
\end{itemize}

\noindent An important suggestion for experimentalists is that, in view of these findings, the studies of the symmetry energy through isoscaling must be focused at colliding energies leading to relatively cool systems.  Another recommendation for future work is the need to decouple theoretically the symmentropic and nuclear contributions to the isoscaling parameters; we are currently working on such a problem.

\section{Acknowledgments}
C.O.D. is a member of the "Carrera del Investigador" CONICET. C.O.D. acknowledges the support of a grant from the Universidad de Buenos Aires, CONICET through grant PIP5969.

\end{document}